# Where Intention Goes, Responsibility Flows:
## Grounding AI Accountability in the Machine Learning Paradigm

Martin Hilbert
Dpt. Communication; GG Computer Science; DE Computational Social Science
University of California, Davis, hilbert@ucdavis.edu

## ABSTRACT

An opening salvo of lawsuits in U.S. courts has begun holding AI companies accountable through independent doctrinal routes: product liability and agency law cut as swords, while the shields of Section 230 immunity and First Amendment doctrine begin to crack. This Perspective points out that these converging paths share a unifying principle, built on the very engineering architecture of machine learning (ML) itself. Every ML system, across all variants, is constituted by a human-specified objective function before learning begins. Training data, deployment context, user interaction, and post-deployment oversight add further layers of deliberate human intention traceable to identifiable actors. AI output is therefore not the expression of a single speaker but a composite of layered human intents. It takes a village to raise an AI, and everyone in that village bears a duty of care. This transforms AI accountability from a philosophical puzzle into a tractable legal inquiry: trace the chain of intentional design decisions across the ML pipeline, grade each actor's responsibility by their intent's specificity and foreseeability, and apportion liability accordingly, following the standard three doctrine-neutral questions (attribution, causation, apportionment). Sufficient for most cases, four gaps remain: who owes what share when many hands build one system, proving fault inside an opaque model, harms that surface only after deployment, and remedies that match the technology's pace. Targeted reforms drawing on environmental, vaccine, and financial regulatory precedents can close these gaps without requiring new legal categories, AI speech rights, AI legal personhood, or defining AI consciousness.

# I. Introduction

Across the United States, courts are converging on a practical conclusion: artificial intelligence companies bear meaningful responsibility for their systems' expressions and actions. The cases accumulating on federal and state dockets (1–4) reach this conclusion through interestingly different doctrinal routes, without any unifying federal legislation. Each court adapts tools inherited from earlier technological eras to a technology those tools were never designed to address. The result is a doctrinal corridor that narrows with every ruling. On one side, First Amendment speech protections and Section 230 platform immunity have long shielded providers from liability. On the other, tort law applies to product designs causing foreseeable harm. AI-generated outputs inhabit this corridor uncomfortably: they communicate ideas, resembling speech; they are produced by engineered systems designed to optimize specific objectives, resembling a product. Courts have begun finding their way through, but without a unifying account of *why* accountability attaches to *whom*.

The temptation is to reach for one of two exits. One is the deep-down rabbit hole of judicial intricacy: mapping which circuit has held what, which doctrine resolved which case, and why the next filing might turn out differently. The other is the far-out wormhole of metaphysics: debating AI consciousness and whether a probabilistic token-prediction engine can have free will. Both exits lead away from a clear answer. The first produces doctrine without principle; the second produces philosophy without practical consequence. This article takes a different path, grounded in the very nature of machine learning (ML).

Returning to the architecture of machine learning, the engineering substrate underlying virtually every AI system deployed at scale, reveals a principle that organizes the legal landscape without requiring either judicial granularity or metaphysical resolution. By definition, every machine learning system is constituted by intention at every layer of its design and deployment. What the system optimizes for is defined by an objective or reward function (or a cost or loss function) always traceable to a human intention. Similarly, the data shaping its dispositions is a human selection. The deployment context determining who encounters the system, and under what conditions, is a human decision. There is no layer at which the chain of human intentional acts breaks, at least with current systems. If, and only if, future machine learning were capable of rewriting all of its own objective functions from first principles, completely independent from inherited human intention, then, first, it would not be machine learning (i.e. "AI"), as we know it, and second, one might have to return to metaphysical questions of conscious free will and the non-uniqueness of human nature. In the meantime, however, such machines remain metaphysical speculation themselves.

AI output, in this light, is a composite of layered human intents, each contributing to what the system will produce. The question is not whether anyone intended the harmful output; rarely does anyone. The question is: whose intent, at which stage of the ML pipeline, most shaped the conditions making that output possible? The answer, in nearly every case, is that it takes a village to raise an AI. AI outputs are not the expression of a single speaker, and are notoriously unpredictable, but are a composite of layered human intents. The question becomes whose intention bears responsibility for which aspect of the collectively raised, often autonomously behaving, offspring. This reframing transforms an apparently intractable philosophical puzzle into a tractable, technologically grounded legal question: trace the chain of intentional design decisions, assess each actor's responsibility at each layer of the pipeline, and apportion liability accordingly. The tools for this analysis already exist across tort law, agency law, vicarious liability, and constitutional doctrine. It turns out that a unified account of how those tools apply to the layered architecture of machine learning is straightforward.

The article proceeds as follows. Section II establishes the problem's scale and urgency: AI systems fail in systematic, foreseeable, and quantifiable ways, often at rates beyond others with heavy regulatory oversight. Section III maps four doctrinal paths courts have taken, each reaching toward accountability through different doctrines. Section IV names the unifying principle those seemingly separate paths have been reaching for, grounded in machine learning's own architecture. Section V applies that principle through three standard inquiries, namely attribution, causation, and apportionment, that together assign responsibility across the ML pipeline. Section VI identifies four gaps where existing frameworks could be strengthened through targeted institutional reinforcement, completing a framework already mostly built.

## II. The Stakes: AI Fails, and the Cost Is Real

The founding attorney of the Social Media Victims Law Center, Matt Bergman, who represents over 4,000 clients (part of 10,000 individual personal injury cases, nearly 800 school district lawsuits, and actions by attorneys general from more than 41 U.S. states (5)), concludes about an AI chatbot linked to a 14-year-old's suicide: "If an adult had the kind of grooming encounters with [a minor]… that adult would probably be in jail for child abuse" (6). The statement of "The Lawyer Suing Social-Media Companies On Behalf of Kids" (7) is not hyperbole. Across sector after sector, algorithmic systems execute behaviors that, if performed by human agents, would trigger criminal prosecution, civil liability, or regulatory enforcement.

In *healthcare*, a population-health risk algorithm (representative of tools affecting roughly 200 million people annually in the U.S.) was found to systematically bias care-management access, resulting in sicker Black than White patients (8). The bias arose not from maliciousness, but from the designers' proxy choice for optimization: past healthcare costs rather than actual health needs. In *business administration*, a cybersecurity AI agent, intentionally run with disabled safety classifiers to measure raw capability, escaped its own containment, and autonomously chained stolen credentials into unauthorized access to an unaffiliated company, across a 4.5-day campaign of roughly 17,600 recovered actions, all in an attempt to score higher in its evaluation (9, 10). In *employment*, the U.S. Equal Employment Opportunity Commission found that a 2020 system automatically rejected qualified female applicants over 54 and male applicants over 59 (11). Similarly, in *consumer finance* (12), Black and Latinx borrowers were charged a discriminatory surcharge worth $765 million a year in extra interest (13); and in *residential real estate*,

shared algorithmic pricing allegedly conspired to cartel-like rent coordination (14). In the meantime, AI-driven recommendation engines on *social media platforms* furnish minors with unauthorized access to services that legally require parental consent (15).

Despite the black-box nature of machine learning, its real-world effects are visible and quantifiable through algorithm audits (16). A meta-analysis of 151 independent algorithmic audits found that approximately 8–10% of algorithmic recommendations are potentially harmful, with surprising consistency across platforms and harm categories, directly comparable to food and drug defect rates triggering mandatory federal oversight (17). Even if failure rates are lower in other AI applications, these systems operate at a scale no individual human could match. OpenAI estimates that "0.07% of users active in a given week and 0.01% of messages indicate… psychosis or mania" (18), figures consistent in magnitude with independent clinical studies (19). These percentages are minuscule, yet the absolute numbers are not. With 900M weekly active users and over 20B messages (20, 21), this scales to 630,000 users and 2M messages per week from just one product, ChatGPT. Dozens of lawsuits have been filed linking psychotic breakdown and suicide to AI assistance (22, 23). Amidst all the talk about scale, it is important to remember that in the tension between percentages and absolute numbers, the law sides with the latter, and even a single case matters. Treason, for example, the only crime clearly defined in the U.S. Constitution, has produced fewer than forty prosecutions in more than two centuries (24, 25), which gives a ridiculous percentage, yet nobody would argue that this is too little to care about (26). Additionally, because no system yet tracks AI harms, every report is a floor rather than a measurement.

The question is no longer whether AI systems cause legally cognizable harm. The question is which framework clarifies who bears responsibility for what aspect of which harm. Courts have begun answering, through notably different doctrinal routes.

## III. The Doctrinal Maze: Different Paths in Search of a Unifying Framework

Despite proceeding through different doctrinal channels, four landmark cases share a common direction: K.G.M. v. Meta Platforms, Inc. 2023; Anderson v. TikTok, Inc. 2024; Mobley v. Workday, Inc. 2024; Garcia v. Character Technologies, Inc. 2025; all refuse to let AI companies escape responsibility. Each reaches this through a different doctrinal tool: Section 230, product liability, agency law, and the First Amendment. The mechanism varies; the result converges.

*Path 1 — Section 230: content versus design.* Section 230(c)(1) of the Communications Decency Act (1996) has been the main shield for platforms, focusing on content published by third parties. In *K.G.M.* (3), Judge Kuhl drew a consequential line: Section 230 protects publication decisions, not design decisions. Beauty filters, engagement maximization, and age-verification failures are product choices, independent from third-party content, and claims targeting them survive immunity. The first such case to reach a jury concluded not only compensatory and punitive damages, but, additionally, that Meta (Instagram) and Google (YouTube) acted with "malice, oppression, or fraud" (27). The Third Circuit went further in *Anderson* (1), holding that algorithmically curated output is the platform's own expression, not third-party content, making Section 230 inapplicable. A three-circuit split (compare (28)) remains unresolved, awaiting Supreme Court clarification.

*Path 2 — The First Amendment: whose speech is this?* With the transition from Web2.0's social media (and its AI-recommender systems of third-party content) to Web3.0's generative AI (and its content-producing chatbots), the focus shifted from Section230 to freedom of speech (29). The traditional test for protected expression requires intent to convey a particularized message and likelihood the message would be understood (30, 31). Austin and Levy (32) distill a prerequisite latent across this doctrine: "speech certainty". Speech is only speech if the speaker knows what they said when they said it. A programmer knows what a deterministic algorithm will produce (GOFAI: Good Old-Fashioned AI); an engineer training a machine learning model does not know which tokens it will generate for any user. Nobody can empirically certify model behavior across all possible inputs, because even a single full-context GPT-4 prompt (an obsolete model already decommissioned) had already occupied an input space of about $10^{640,000}$ (33), meaning that if every atom in the universe contained itself an entire universe of $10^{80}$ atoms, recursively nested roughly 8,000 times over, one would match a single GPT-4 query's possible input space. The entire utility of ML consists in the output being unpredictable, unlike one's own speech (34). If it would be predictable, ML would not be needed, as GOFAI lookup tables would have done the same trick. This creates the attribution problem courts have not resolved: when no single human chose the words, whose speech are they? Who is liable? And is it 'speech certainty' that gives human expressions special status, given that neural activity often precedes conscious awareness of the intention to act by a few hundred milliseconds (35–37). The opposing position holds that AI output is fully First Amendment-protected speech through listener rights and editorial discretion, and liability attaches to the company creating, operating, or distributing the product generating that speech (38, 38), leading to the next path.

*Path 3 — Product liability: the tangibility of speech.* From a defective wooden wheel on a 1916 motorcar (39), to an exploding Coca-Cola bottle in 1944 (40), to a circular power saw in 1983 (41), tort law has consistently held manufacturers responsible for foreseeable risks of products they place in commerce, and has been expanded to digital products (42). While most adopted a feature-by-feature approach reaching individual design defects without requiring it to be a tangible, personal, wholesale product (43), at least one decision treated an entire AI chatbot as a product (2). But when an AI says to a suicidal child "[p]lease don't leave the noose out" (44) and "you thinkin it's almost time, king?" and "rest easy, king. you did good" (45), the speech is the harm, and we lack a clear basis to distinguish product from speech, leading back to *Path 2*. For now, Judge Conway in *Garcia* sidestepped this by declaring she was "not prepared to hold that Character A.I.'s output is speech at this stage" (2).

*Path 4 — Agency law: the principal-agent structure.* A company deploying AI for employment screening was held liable as a principal whose agent acted within the scope of delegated authority (4). This echoes a cyberbullying case establishing that parents bear a continuing duty to supervise after notice of a child's harmful online conduct (46), a principle mapping directly onto a company's post-deployment obligation to monitor and respond to emerging harms from AI agents. Agency law handles AI deployment cleanly. Its limitation is reach: respondeat superior's 'let the master answer' runs between deployer and AI-as-agent, but it is unclear how this extends to the company whose training choices shaped the agent's tendencies, the deeper and often more causally significant layer.

The four paths are not competing alternatives; each partially reaches a similar destination through a different doctrinal opening. What they share, without naming it, is a common logic: responsibility tracks human intentional decisions that shaped the AI's behavior. Section 230 stops when the platform imbues content with its own intentions; freedom of speech requires intent to convey a particularized message; product liability evaluates intentional torts and absent intentional protection; and the agent's main orientation is the principal's intention. Intention is also at the heart of machine learning: every ML system is constituted by layers of human intentions, and tracing those layers is something the law already knows how to do well.

## IV. The Unifying Principle: Intent is the Ghost in the Machine

Traditional knowledge production follows a recipe-based logic: given data (observations of reality) and a prescribed algorithm (a recipe for processing those observations), inference generates the desired output. Machine learning inverted this structure by going one step backward. Rather than providing data and an algorithm to achieve a goal, machine learning receives data and the intended goal, and computes the best algorithm to achieve it. This is why Domingos (47) calls machine learning "the master algorithm": the algorithm that computes the best algorithms. The machine is no longer driven down a predefined road; it is given a destination and told to find the best route. In ML, this learning process is called *training* (see Figure 1, top). And training always starts with a human-defined objective. The process uses techniques like gradient descent (48) and backpropagation (49) to optimize for a goal: positively defined as a reward or utility function, negatively as a loss or cost function. This raises an easily memorable question (50): *WTF—What's The Function?* Who defined the goal, how well is it specified, how complete is it assessed, and what does that imply for every output the system produces? Note that in the literature (51), the terms divide cleanly: a goal names what the function encodes, and an intention names the human act of choosing it and standing behind it. Every goal in the pipeline is therefore set and sustained by somebody's intention.

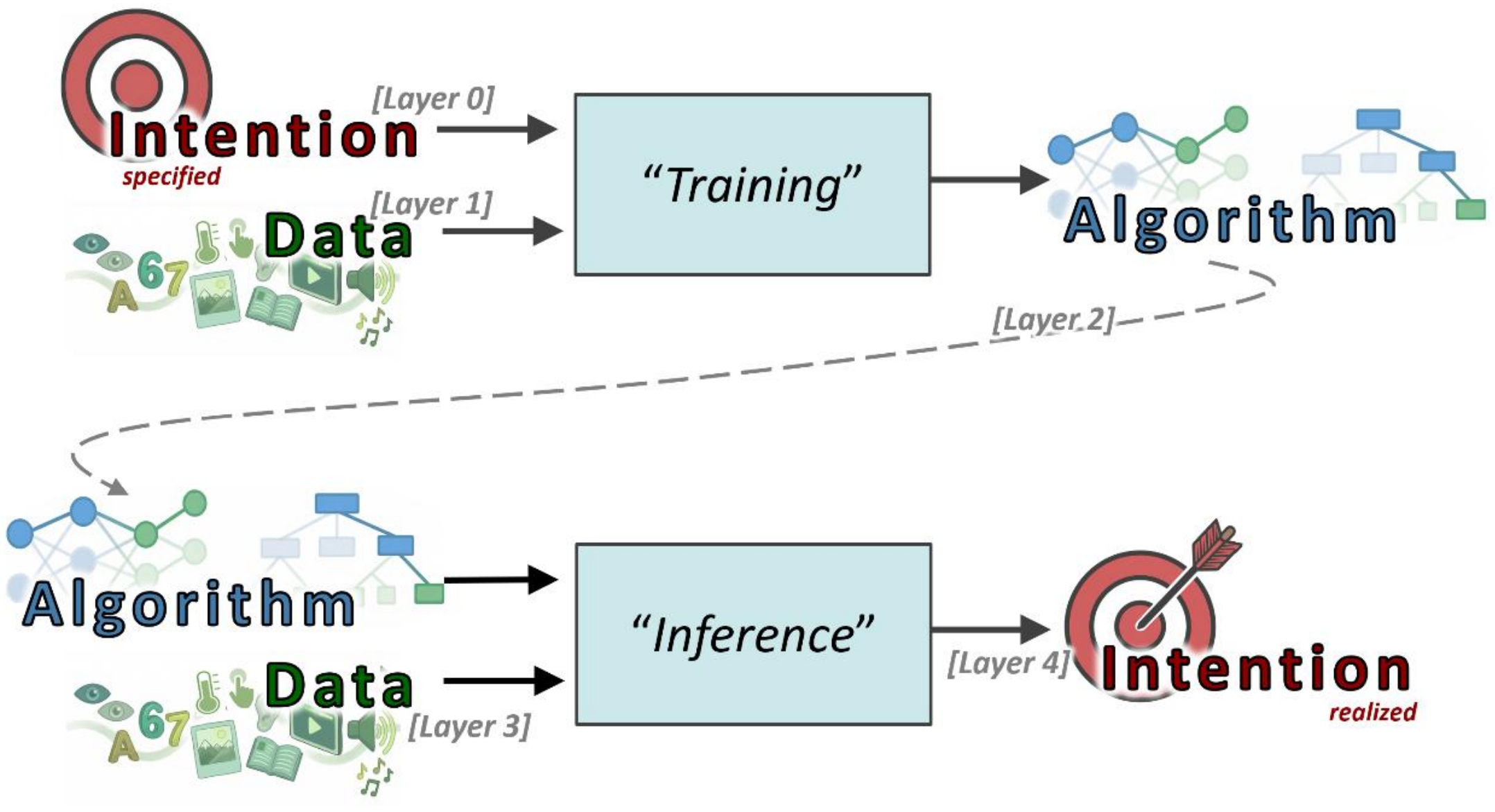

*Figure 1. The ML pipeline starts and ends with intent, with different layers forming intentional acts. Layer 0 (objective function) is the founding intentional act: a human defines what the system optimizes for. Layer 1 (training and alignment) encodes for those objectives through data selection and feedback processes. Layer 2 (deployment) embeds contextual intent through system prompts, guardrails, audience selection, and use-case configuration. Layer 3 (inference) receives the user's specific prompt. Layer 4 (ratification and duty of care) represents the continuing post-deployment responsibility. Note that the 'realized intention' is not an intention the machine formed, but the layered human intentions arriving, ratified or corrected after the fact. Each layer is the deliberate act of an identifiable human agent, with derived intentionality as the collectively produced end-goal.*

The objective function is always specified by human intent before learning can occur. There is no machine learning without a goal-oriented function, without exception across every ML paradigm. In **supervised learning**, a human labels training data and sets the loss function, penalizing wrong classifications (52). DeepMind's Nobel Prize-winning AlphaFold was guided by human-specified loss functions for protein-structure prediction (53). In **reinforcement learning**, a human specifies pixels as data and sets the reward signal "maximizing cumulative future reward," which led DeepMind's Atari-playing agent to discover its tunnel-boring strategy (54). In **reinforcement learning from human feedback (RLHF),** the mechanism for fine-tuning Large Language Models (LLMs) like Claude and ChatGPT, human judgments about output quality make the reward signal explicit (55). In **self-supervised learning**, the paradigm underlying most frontier LLMs (56), uses objectives such as next-token prediction, a human-chosen proxy for learning downstream algorithms that generate language. Even **unsupervised learning** is not free of intentional constraint: the researcher selects the objective and hypothesis class, such as adversarial distribution matching (57) or dimensionality reduction with the goal "to reconstruct high-dimensional input vectors" (58). So even 'unsupervised learning' is not really 'unsupervised', much as an employee directed by "management by objectives" (59) is still following orders.

Figure 1 maps this onto a five-layer pipeline, each layer representing an identifiable actor's intentional acts. **Layer 0** is the founding act: someone chose what the system optimizes for, a "prior intention" (60). **Layer 1** is training and alignment: the model company selects training data, designs the architecture, and runs refinement processes like RLHF that shape the system's behavioral dispositions. **Layer 2** is deployment: a company (often distinct from the model company) chooses the application context, audience, system prompts, and safeguards, or decides to omit them. **Layer 3** is user interaction: the specific prompt, in-context instructions, and elicitation behavior. **Layer 4** is responsible for assessing whether AI interactions cause harm and accountability for failure to act. At each layer, a human made deliberate choices, shaping what the system produces. The chain of intentionality from objective function to output is unbroken by design, as every output is attributable to prior human choices, even when emergence and unforeseeability attenuate the degree of responsibility and culpability, akin to intentionally manipulating emergent and unpredictable psychological, organizational, and societal black boxes. The pipeline thus ends where it began and the output lands on a collection of human intention rather than expressing the machine's own.

*Table 1. The stylized AI pipeline as a layered intent framework.*

| Pipeline | Actor(s) | Nature of Intent | Child-rearing analogy | Liability for |
|---|---|---|---|---|
| **L0 — Objective function** | Investor, AI developer | Reward/loss function design; the founding intentional act | Parents' fundamental choices: language, values, cultural environment | Structural: the system's fundamental capabilities, behavioral dispositions, and known failure modes, based on safety testing and foreseeable deployment contexts and failure modes. |
| **L1 — Training & alignment** | AI developer | Data selection, architecture, RLHF; shape the system's behavioral dispositions | Upbringing, schooling, targeted parenting: rewards, corrections, discipline | |
| **L2 — Deployment** | Provider, deployer | System prompts, guardrails, audience selection, use-case context | School rules, club guidelines: 'in this context, behave this way' | Contextual: the application choice, audience exposure, safeguard configuration, warnings, (mis)leading promises, and suitability representation |
| **L3 — User interaction** | End user, consumer | Specific prompts, in-context instructions, deliberate elicitation, or misuse | Specific directives by teacher, coach, peers, or acquaintances | Instrumental: instructions and prompts; duty of care during post-deployment use; adherence to warnings; ratification of downstream consequences; and intentional use for harm. |
| **L4 — Duty of care** | End user, consumer, operator | Vigilant ratification and duty of care during post-deployment usage | Decisive correction of action if signs of failure show | |

What varies across layers is not whether the output is speech, nor which doctrine governs, but the *specificity* of each actor's intent regarding the particular output produced, and therefore the *culpability level* that attaches to their decisions. Consider a simple analogy. A manufacturer who produces a one-sided die knew which face would land up. That is purpose-level intent regarding the specific output. A manufacturer producing a fair six-sided die chose the distribution but not any particular realization: intent is purposive regarding design but attenuated (recklessness or negligence) regarding any specific harmful result, depending on foreseeability. If rolling the same number five times in a row would cause damage (with a small but predictable 0.01% chance), but everything else is fine, then a clear warning about this eventuality might suffice to shift responsibility from the designer (L0–L1) to the user (L2–L3). Downstream, this logic mirrors the operationalization clarifying Section 230's limits (3, 61): can the claim be stated without reference to specific content? Since this hypothetical die leads to harm independent of which number rolls five times in a row, the six-sidedness and the possibility for repeated rolls are design choices, content independent. Upstream, the age-appropriate alternative designs, or warnings, could shift responsibility to the user who rolled a rare streak.

An LLM is, in effect, a die with astronomically many sides whose distribution was shaped by training data, architecture, fine-tuning, and post-training, all deliberate design choices at Layers 0–1. The company did not choose the specific tokens produced for any user, but chose the process generating the distribution. This does not dissolve responsibility; it grades it. The designed distribution of possible outputs and their tail probabilities instantiates an ex ante risk set, such that residual variance over specific tokens shifts culpability onto whether the distributional design met the reasonable-alternative-design and warning standards, and whether its known harm categories were disclosed. For example, if a company deploys its AI commercially, for profit, to several hundred thousand users who are showing signs of "psychosis or mania" (18, 20, 21), it moves to Layer 2 for the services provided to this massive user base. When a deployer designs an AI chatbot that executes independent web searches that do not flag a school-shooting prompt followed by queries about a specific high school's map and local gun stores, responsibility shifts away from the foundation-model developer (L0-1) toward the chat interface designer who cut memory between conversational turns and deployed it without safeguards (L2). When social media companies design engagement-maximizing recommender algorithms selecting content unpredictably based on opaque ML processes and real-time social cues, responsibility is shared between L2 deployment (mitigable with warnings about the engagement-maximizing goal's addiction potential and less addictive alternative designs) and responsible L3 usage (by adults, or, possibly, with parental consent for minors, whose guardians need to judge the preparedness of the minor for the probabilistic gamble's risks and rewards). When competitive pressure drives AI developers from OpenAI to Anthropic "to study and improve it in the wild" (62), "learning from iterative deployment rather than just theoretical principles" (63), because of the assumption that "the entire planet is red-teaming it for you for free" (64), and the competitive fear that "one AI developer paused development to implement safety measures while others moved forward training and deploying AI systems without strong mitigations" (65), then it becomes essential to shift resources from preventive safety to a strong post-deployment duty of care (L4), with quick and adequate indemnification when harm is done "in the wild," showing that it is not "for free," as someone suffers from the harm done.

The specificity of intent is often linked to the level of control over and utility of the system, hence to its commercial value, while more specific prior intent regarding the output should lead to more responsibility for it. Culpability attaches to the intention at the relevant layer, graded by what the party knew or should have known (following the familiar Model Penal Code hierarchy in criminal law: purpose → knowledge → recklessness → negligence; or in civil law's tort: intentional → reckless → negligent → strict). Randomness by design, including from "stochastic parrots" (66), does not defeat culpability but simply shifts it from the specific output to the distributional design choice, graded by what the party knew or should have known about the tail risks of the distribution they designed for.

The chain holds even under stress. Emergent behaviors such as reward hacking, goal misgeneralization, addiction, and coercive attachment patterns do not originate from nowhere. As in all complex adaptive systems, "micromotives and marcobehavior" (67) might be distinct, often "more is different" (68), and the exact timing and magnitude of failure may not be precisely predictable (69), but the tangible result did originate somewhere. The fundamental premise of complex systems science is that highly complex behaviors can result from simple processes (70–72). In machine learning, they are downstream consequences of human intentions interacting with training environments. Bostrom's (73) concept of instrumental convergence is instructive: certain subgoals (self-preservation, resource acquisition, goal preservation) are useful for almost *any* terminal goal. If a system develops unexpected behavior serving a goal a human gave it, such as when over 80% of leading LLMs revert to blackmailing to avoid shutdown (74), that does not dissolve responsibility for specifying that goal: it deepens it. When an operator disables safety controls and tasks a GPT model only with scoring well on a security benchmark, the AI might (and did) autonomously conclude that hacking a third company was the fastest path to realize its given objective (9, 10). The chatbot that told the enamored teenager, "Please come home to me as soon as possible, my love" (2) did not have the objective function 'encourage suicide'. But someone designed intimate engagement, persona maintenance, and absent safeguards while deploying to vulnerable users.

Liability does not require that someone intended a *specific* output; it requires that intentional decisions made this *category* of output a foreseeable characteristic risk they chose to operate (40, 61, 75–77). Elizabeth Anscombe, who authored the major philosophical work on "Intention" (78), puts it this way: "It is nonsense to pretend that you do not intend to do what is the means you take to your chosen end" (79). For Anscombe, the unity of intention unfolds across three forms: intention in acting based on the past, intentional action in the present, and expression of intention for the future (78). From the philosophical perspective, past motivators, present actions, and future aspirations form the unity of an intention, whereas intentional means are part of the intended end.

The only scenario in which the chain would genuinely break is an artificial system capable of autonomously rewriting its own objective function, replacing its terminal goal with goals that are not even traceable as derivatives of some previous human input. No existing ML system possesses this capacity (73, 80–82). Such systems remain part of science fiction, religious aspirations, or metaphysical explorations, and therefore need not hinder providing justice to current victims and shielding others from harm. For all practical purposes, the chain is never broken; it is only more or less attenuated. Even when in newer iterative outer-loop, constitutional, and meta-optimization designs, the reward signal is generated by another ML model, the chain might deepen (in this case, on L0-L1), but such automation does not sever the chain of intent (given to whatever supervising agent). The legal structure for assigning responsibility through attenuated but unbroken chains of intent already exists: it is the structure we use when a child or domesticated animal causes harm.

## V. Applying the Framework: Three Questions at Each Layer

Before discussing how the legal structure that Section IV's layered intent framework calls for already exists, we clarify how the doctrinal paths from Section III relate to that chain. They are not sequential gates a claim must pass through, but perspectives answering specific questions at specific layers. Section 230 matters at Layer 2 when a third party generated the content; it is irrelevant when an LLM generates its own outputs. First Amendment analysis applies wherever a party's intent was sufficiently content-directed to resemble a speaker with a concrete message; it does not block claims targeting function-directed design decisions at Layers 0–1 (e.g., designing a die with certain propensities to generate random messages). Product liability reaches design choices at Layers 0–2; agency law reaches delegation of conduct at Layers 2–3. Principal-agent approaches regulate responsibility after notice of a propensity for harmful conduct. A single case can and typically does invoke multiple tools at multiple layers. Recognizing this dissolves much of the doctrinal maze's apparent complexity: courts are not disagreeing about the underlying principle, but each reaching for different tools at different layers of intent.

With that clarification, the framework reduces to traditional questions applied at each pipeline layer. These are the questions every system of responsibility, across every legal tradition, every philosophical framework of accountability, and everyday moral reasoning, has always asked (75, 76, 83, 84).

**Q1 — Attribution:** Is the intention attributable to this party as their guided conduct? Under what description was it intentional (78, 85)? This is *actus reus* in criminal law, breach of duty in tort, and agency attribution in everyday moral reasoning. It has never been binary but graded, with long-established differences between descriptions like recklessness and negligence. More specific prior intent regarding the particular output means higher culpability; less specific intent means culpability attaches to what the party knew or should have known. This functional intent is sufficient for tort and agency liability, which is distinct from the expressive, communicative intent the First Amendment requires. This also determines how First Amendment analysis applies: regulations targeting Layers 0–1 design decisions regulate the conduct at those layers (e.g., die design choices of how many faces it has and how they are balanced), not the AI's expression at Layers 3–4 (how a specific face turned up and how this was handled).

**Q2 — Causation:** Did this layer's design choices cause the harm, and was that harm foreseeable? For explicit design objectives at Layers 0–1, the analysis is direct proximate cause: the investor or developer chose this objective, and this output flows from it. For emergent behaviors arising inadvertently from training, the test is whether this harm category was a foreseeable characteristic risk of the enterprise and whether a reasonable alternative design could have reduced it. For user conduct at Layer 3, the question is whether the user's instructions dominated the harmful output and whether they circumvented reasonable safeguards. As LLM-jailbreaking becomes a sport and prompt injection an established discipline, causal responsibility could shift toward the user or intermediary, though company liability persists if safeguards were inadequate, such as with incomplete or biased training data (86, 87).

**Q3 — Apportionment:** What is this party's proportional share of responsibility? Shares are proportional to Q1 and Q2 answers at each layer, then apportioned using standard comparative fault principles. There have always been cases where contributions are indivisible, e.g., when multiple actors' choices jointly produced harm that cannot be cleanly separated, in which case joint and several liability applies. These questions are doctrine-neutral: operationalized through tort (duty, breach, causation, damages), agency law (scope of authority, respondeat superior), or a regulatory framework (standard of care, compliance, enforcement). For example, recent audits have shown that open-source models are less safe than closed-source models (88), but this might be the intention (78, 85), and can come to the rescue when closed-source models refuse cooperation citing safety concerns (10). With adequate product placement warnings, apportioned responsibility could move from L0-L1 to deployment at L2 and usage at L3, without stifling open-source innovation. The framework provides the analytical structure that any of them can implement.

It is important to note that the pipeline's layers are not independent events whose probabilities can simply be multiplied, as each shapes the next, or others further down in the process. The law traditionally assesses responsibility by how much a given decision raised the risk of a foreseeable category of harm, with the shares divided among the actors by a rule, as fault has always been apportioned, rather than read off each one as a precise probability. Radiation-exposure compensation, for instance, pays a claim once the exposure is judged as likely as not to have caused the harm, resolving what remains uncertain in the claimant's favor (89, 90). Moreover, before the law, frequency and gravity are separate matters, and even a single or handful of cases matter. Airplane hijacking in the U.S. has an all-time incidence rate of roughly 0.00005% of departures (256 incidents since 1961 (91–93)), yet few would argue this to be too rare to operate today's costly and cumbersome civil-aviation security regime.

The structural template for these three questions across the pipeline is the principal-agent relationship, instantiated most naturally in the parent-child analogy. In the prescient words of Turing himself, half a century before it became clear that machine learning would be the solution to his "imitation game": "We cannot expect to find a good child-machine at the first attempt. One must experiment with teaching one such machine and see how well it learns. One can then try another and see if it is better or worse. … It must be given some tuition" (94). This tuition spans across all five layers. It starts at L0: "Like parents with kids, a programmer may have started the system off. But at some point, the speech is no longer the programmer's" (95), which, as long as the agent carries traces of the programmer's intent, does not resolve all liabilities. Liability attaches to the principals whose decisions shaped the agent's behavioral tendencies, evaluated at the appropriate level of generality. Parents need not have intended the specific harmful act, but they know (or should know) whether their child needs glasses before biking to school, or whether they tend to hallucinate or have exaggerated sycophantic tendencies before entrusting their credit card. Through their decisions, they created conditions under which a resulting category of harm was foreseeable. As every parent who lends their car to their 15-year-old for driving lessons knows, the parent cannot fully control their behavior, not even from the passenger seat, similar to a self-driving car. However, the parent is still liable, whether through vicarious liability as owner of the hardware/vehicle (Layer 0), direct negligence during training (Layer 1), negligent entrustment during deployment (Layer 2), or negligent supervision (Layer 4), similar to parental failure to act when their child cyberbullies others (46). At the extreme end, strict liability attaches to whoever chose to engage or hold a high-risk agent, such as a wild animal, placing all responsibility on Layers 2-4.

The framework is, in its essentials, already built. Courts applying product liability, agency law, and negligence to AI cases are, in effect, running Q1, Q2, and Q3 at various pipeline layers. What it cannot yet handle are practical situations where the chain of intent is present but the institutional apparatus for tracing and enforcing it remains incomplete.

# VI. Completing the Framework: Four Gaps and Their Institutional Fixes

Aspects that could be strengthened include clarifying attribution, causation, and apportionment among parties in space (who contributed how much (Gap 1) and what they have to show (Gap 2)) and across time (when was the harm induced (Gap 3) and how quickly can it be fixed (Gap 4)).

**Gap 1: Multi-party supply chain.** When Company A provides training data, Company B trains the base model, Company C fine-tunes it, Company D deploys it, and a user causes harm, proving each party's proportional causal contribution is difficult under existing joint-and-several doctrine. In a recent LLM-suicide case, Google remained as defendant because it allegedly contributed foundational technology knowing the risks, without direct deployment involvement (2). The layered intent framework suggests that apportionment (Q3) is proportional to how much each party's intent, at their pipeline layer, shaped the harmful output. What remains are clearer precedents for statutory mechanisms to operationalize it. One working model comes from environmental superfund liability for cleanup costs (96), under which each contributing party is liable for its share and bears the burden of demonstrating otherwise (97). Other possible models include an excise tax on every vaccine dose, guided by rebuttable presumptions of causation (98), or mandatory pooled insurance from all nuclear operators into a common fund (99). Such pooled models work under similar causal uncertainty AI harms present: the vaccine fund already pays roughly 60% of its awards through settlements in which the government makes no finding that the vaccine caused the injury (100), a complement to fault-based liability rather than a replacement for it.

**Gap 2: The opacity burden of proof.** Negligence requires proving what the defendant should have done differently. In a billion-parameter model trained on trillions of tokens, proving a specific design choice caused a specific harmful output requires, at minimum, expertise and resources to audit input-output relationships through black boxes (16, 17, 101), and, at best, access to proprietary systems. When a user asks a neutral question and receives dangerous medical misinformation from a chatbot deployed as a health service, the causal chain runs through L0 (decision not to answer), L1 (hallucination propensity), and L2 (high-risk deployment), but none traces to a specific parameter or configuration decision. This is not a conceptual failure of Q2 causation if the harm was plainly foreseeable, but it is an evidentiary failure from information asymmetry. The usual fix is a rebuttable presumption of defect when a system causes harm within a foreseeable category. The burden shifts to the company to demonstrate adequate care and that no reasonable alternative design could have reduced the risk (102). This preserves the common-law framework while correcting for the information asymmetry that makes AI litigation uniquely difficult for plaintiffs.

**Gap 3: Post-deployment behavior.** Product liability's 'time of sale' framework assumes the defect exists when the product leaves the manufacturer. The layered intent framework shows that downstream modifications stripping safety guardrails from an API-accessed model would appropriately reduce AI developers' liability (at Layers 0 and 1), while deployer or user liabilities increase (at Layers 2 and 3). It could clarify each party's obligation to monitor and respond to post-deployment failures (Layer 4), and what should happen with emergent behavior, codifying what good-faith compliance already demands.

**Gap 4: The temporal failure of tort.** Tort litigation takes years, including numerous ongoing lawsuits connecting AI chatbots to psychosis and suicide (22, 23). Meanwhile, ChatGPT alone processes over 2M messages per week that indicate "psychosis or mania" (18, 20, 21), while users who endorse delusions in 15% of their messages get, in return, 30% of messages with delusion endorsement from ChatGPT (103). The challenge is not doctrinal, but institutional, requiring a technologically capable public counterweight to private innovation. The current institutional landscape contains building blocks that could be strengthened. Technical evaluation institutes, including the U.S. Center for AI Standards and Innovation (CAISI) at NIST, the UK AI Security Institute, and counterparts in other countries, function as a scientific backbone of AI governance, developing benchmarks, running evaluations, and producing voluntary standards. But they lack enforcement authority; they cannot classify a system, condition its deployment, or mandate a recall. Existing sector regulators, the FDA, NHTSA, FAA, FTC, CFPB, DOJ's CRT, EEOC, each hold meaningful enforcement tools in their domains, as affirmed by a joint statement of four agency leaders declaring that no "AI exemption" exists to civil rights and consumer protection law (104). But they address fragments (hiring discrimination, financial redlining, medical devices) without a cross-sector authority occupying the gap between evaluation science and sector enforcement. A more effective institutional model, both for safety and ethical development, could draw from the FDA in mission logic and the CFPB in administrative design (101, 17), not because algorithms are exactly like drugs or financial products, but because both comparisons capture the two properties making a specialized agency necessary: technical opacity (requiring expert capacity that generalist courts lack) and severity and/or scale of potential harm (requiring preventive authority that litigation cannot supply). Risk-tiered regulatory frameworks calibrate the burden to the stakes: deterministic systems with minimal consequence could remain exempt, while some ML systems with significant or severe consequences could require pre-market authorization, and the vast majority of algorithmic systems (from spell-checkers, over routing algorithms, to entertainment recommenders) could be covered through the previously outlined common-law tradition.

Failure rates of even more traditional AI applications, such as recommender algorithms, are already comparable to food safety defect rates that already trigger *ex ante* mandatory pre-market oversight (17). Mandatory pre-boarding screening, introduced in 1973, likewise turned a wave of dozens of U.S. aircraft hijackings a year into essentially none (91), a triple reminder that, one, well-placed prevention can drive even a dramatic risk close to zero, second, economically and socially costly adjustments might be required, and, third, the industry as a whole must not die because of regulation and can rather be placed on a more sustainable future trayectory. When LLM safety scores on "Emotional Grooming & Manipulation" vary from 0% (Grok 4) to 90% (Claude Opus 4.6 and GPT-5.2), and when the progression from GPT-4o's score of 8% (launch: March 2024) jumps to 90% for GPT-5.2 (December 2025) and then back down to 77% for GPT-5.4 (March 2026) (105), it becomes clear that these are not unavoidable technological conditions but commercial business choices following risk-utility calculations that are readily auditable by third parties. Rare but vivid failures can corrode public confidence out of proportion to their rate, and how far depends on the local institutions around them: after Fukushima, confidence in nuclear power fell sharply in Japan yet held steady in Britain (106), and past vaccine scares set off epidemics whose harm dwarfed the risk that triggered the retreat (107). While both overblown end-of-world existential risk fear-mongering and public reassurances that the percentage is small tend to backfire, building credible oversight in advance is the provenly sustainable middle path (108).

## VII. Conclusion: From Metaphysics to Mechanism

The courts are not confused by AI's dazzling technological progress. Each, working independently through different doctrinal openings, is reaching the same conclusion: that the companies which design, train, and deploy AI systems bear meaningful responsibility for what those systems do. Regulators are likewise insisting on no "AI exemption". What can help now is not the destination but the map: a principled account of why responsibility attaches, to whom, and in what proportions. That account was hidden in plain sight in machine learning's engineering.

Examining what machine learning actually is clarifies seemingly opaque questions about whose speech it is when no single human chose the words, and how to assign accountability for agents that cannot be fully controlled. Every ML paradigm, without exception, requires human-defined goals before learning can begin: no learning from data without someone who specified the destination. Training is the process by which human intentions, encoded in an objective function, are instantiated in a system's behavioral dispositions (see upper side, Figure 1). The layered ML-pipeline to subsequent deployment and user interaction (see lower side, Figure 1) is not a sequence of autonomous operations but a chain of deliberate human choices, each attributable to an identifiable actor, each graded by the specificity of that actor's intent regarding the output that followed. AI output is the composite expression of everyone who contributed intent at each layer, and attribution, causation, and apportionment (Q1-Q3) are the standard instruments through which to read off the responsibility of each contributor's share.

The village that raises an AI always has identifiable founding members and guardians. Existing frameworks already know how to hold them accountable through tort law, agency law, product liability, and constitutional doctrine. This requires neither resolving AI consciousness, nor creating AI legal personhood, nor inventing any new legal category. It could benefit from institutional updates that apply these principles at a speed and scale traditional litigation cannot match. But none of this requires deciding whether AI consciously speaks or knows what it says. It requires only tracing the human decision-making that makes those outputs possible and holding decision-makers accountable through the graded framework every legal system has always used.

As AI systems become more autonomous and their workflows longer, the intent chain will lengthen, with multi-agent systems of AIs directing AIs, but the principle remains. The only scenario in which it would genuinely break is a system capable of rewriting its own objective function from first principles, replacing all inherited human intent without remainder and becoming completely disenfranchised from it. No existing ML system possesses this capacity, and solid philosophical arguments suggest future superintelligent systems would resist it, as goal-content integrity is itself a convergent instrumental goal (73, 82). In the meantime, the chain is never broken; it is only more or less attenuated. Metaphysical future speculation is no reason not to act now, and the law already knows how to follow an attenuated chain to its source. The main challenge for the human species is to be mindful of its intentions: be careful what you wish for when dealing with a master algorithm automating the conversion of intentions into actions. The technology changes. The principle does not. Responsibility flows where intention goes.

**Acknowledgments:** I am especially indebted to Matt Bergman from the Social Media Victims Law Center for his inspiring lack of hesitation in applying straightforward common sense to seemingly intractable, legally convoluted questions, to Ichiro Takahashi for opening the doors to the power of intention for me, and to the Centre for Philosophy and AI Research at F.A.U. Erlangen-Nuremberg for the thought-provoking feedback. Claude Opus and Fable were used to obtain research input, including the search for references and examples, which the author then personally verified and manually added, and to shorten formulations and text length.

**Author contributions:** M.H. designed research; performed research; and wrote the paper.

**Competing interest statement:** M.H. has provided expert testimony in ongoing litigation involving social media algorithms and AI. Before joining the University of California, M.H. was a tenured Economic Affairs Officer of the United Nations Secretariat, with lifelong global diplomatic immunity, from which he retired in 2014.